# Features of randomized electric-field assisted domain inversion in lithium tantalate

**Salvatore Stivala,**[1*] **Fabrizio Buccheri,**[1] **Luciano Curcio,**[1] **Roberto L. Oliveri,**[1] **Alessandro C. Busacca,**[1] **and Gaetano Assanto**[2]

[1]*DIEET, University of Palermo, Viale delle Scienze, Bldg. no. 9, 90128 Palermo, Italy*
[2]*Nonlinear Optics and OptoElectronics Lab, University "Roma Tre", Via della Vasca Navale 84, 00146 Rome, Italy*
*\*salvatore.stivala@unipa.it*

**Abstract:** We report on bulk and guided-wave second-harmonic generation via random Quasi-Phase-Matching in Lithium Tantalate. By acquiring the far-field profiles at several wavelengths, we extract statistical information on the distribution of the quadratic nonlinearity as well as its average period, both at the surface and in the bulk of the sample. By investigating the distribution in the two regions we demonstrate a non-invasive approach to the study of poling dynamics.

## 1. Introduction

Electric field assisted periodic poling (PP) is a powerful and versatile technique to achieve Quasi-Phase-Matching (QPM) and efficient frequency conversion with ferroelectric crystals in both 1D and 2D configurations [1–4]. PP-QPM is also crucial towards signal processing via quadratic cascading [5–14], including transverse light localization [15–18], wavelength shifting [19–21], unidirectional optical transmission [22,23]. A few parametric applications, such as backward second-harmonic generation (SHG), counterpropagating optical parametric oscillations and ultraviolet generation, require short periodicities of the nonlinear coefficient distribution, often below 1μm [24–27]. Thus, great attention has been recently paid to small irregularities in the periodic pattern and how they can affect the performance of fabricated devices [28–30]. In fact, the presence of a 2D modulation of the mark-to-space ratio (MTSR) superimposed to a pattern of nominal period Λ (defined by PP electrodes) is inherent to electric field PP and due to the stochastic nature of the nucleation process triggering ferroelectric domain inversion [31].

Randomness in the periodic distribution of the nonlinearity gives rise to several non-collinear grating vectors in Fourier space, thus satisfying momentum conservation in a wide range of wavelengths and yielding broadband SHG [28–30, 32–34]. Such phenomenon, known as random quasi-phase matching (rQPM), is more relevant the shorter the grating period Λ. A non-invasive analysis of the nonlinear susceptibility distribution with its average periodicity along various directions can be crucial in designing and testing parametric components.

Among ferroelectric crystals, Lithium Tantalate (LT) has emerged as one of the most attractive because of its high threshold to photorefractive damage, extended transparency in the UV down to 280 nm [35], large electro-optic and quadratic coefficients [36]. Moreover, LT allows realizing good quality dielectric waveguides by means of proton exchange (PE) [37]; the latter is also compatible with PP and has proven to be an excellent fabrication approach for devices in nonlinear integrated optics [38, 39].

In this paper we discuss bulk and guided-wave SHG via rQPM in Lithium Tantalate, demonstrating how the far-field images of the second harmonic generated at various wavelengths can provide statistical information on the bulk/surface distribution of the quadratic nonlinearity as well as its average period.

## 2. Samples fabrication and nonlinear characterization

We prepared rQPM samples from optical grade wafers of 500μm thick z-cut congruent LT. For the periodic poling we applied high voltage pulses across the LT thickness using an electrolyte gel and an insulating mask. After spin coating the −z facet with a 1.5 μm thick photoresist (Shipley 1813), we defined by standard photolithography a Λ = 1.5μm period grating with grooves parallel to the y-axis. Subsequently, the samples were soft-baked for 30 minutes at 90°C and, after development, post-baked overnight at 90°C and for 3 hours at 120°C. In order to exceed the LT coercive field we used an electrolyte gel over the photoresist insulating mask and applied a single 1.3 kV pulse on a 10 kV bias for 300μs, a time interval long enough to achieve periodic ferroelectric domain inversion in the patterned area of 0.06 x 5.0 mm$^2$. After poling, some of the samples were etched in a solution of Hydrofluoric Acid (HF 40%) and water (HF: $H_2O$ = 1:10) to reveal the domain patterns on the −z surface [40]. Fig. 1 shows two typical scanning electron micrographs of different portions of the sample: while in most of the area the periodicity of the electrodes is faithfully transferred on the sample (Fig. 1(a)), in some portions a random MTSR is clearly visible in the QPM grating (Fig. 1(b)). The poling dynamics is such that the first enucleated (inverted) domains spread in the x-y crystal plane and become wider than those generated later. Despite the dominant role of the periodic electrodes in defining the ferroelectric grating (as demonstrated by first-order QPM SHG on similar substrates [27]), the final MTSR has non-uniform stochastic features superimposed to the regular pattern. While this effect is known to always occur during electric-field poling, it becomes particularly relevant when the domain size (i.e., period) is small, as in our samples.

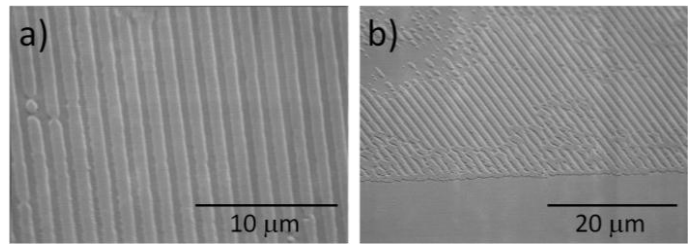

Fig. 1. Scanning electron microphotographs of an etched LT sample after periodic poling: (a) general view and (b) detail of the area by the edge of the poled region.

Slab waveguides were fabricated after poling on the –z facet by "sealed ampoule" proton exchange (PE) [37], using a melt of Benzoic Acid and 3.6% Lithium Benzoate that was proven to preserve the nonlinear optical properties and domain orientation of the crystal [38], realizing "soft PE" waveguides in the α-phase. PE for 144 hours at ≈300°C yielded single mode ($TM_0$) waveguides at the fundamental frequency (FF) wavelengths used in our experiments, with two modes ($TM_0$ and $TM_1$) supported at the corresponding second harmonic (SH) wavelengths. Finally, the chip end-facets were polished to optical grade to allow efficient in and out coupling from/to radiation modes.

For the nonlinear characterization, the FF excitation was provided by a cw Ti:Sapphire laser tunable from $\lambda_{FF}$ = 700 to 980nm, with a 40 GHz linewidth and a maximum operating power of 1W. Bulk measurements were performed by employing an f = 11mm lens to focus the $TEM_{00}$ input beam to a waist of about 27μm at the center of the sample; for the measurements in waveguide an additional cylindrical lens (f = 50mm) was used to obtain an elongated beam of about 27 x 4 μm$^2$ and so ensure a good input coupling into the guiding structure. The measurements were performed with an input power of about 40mW and at the constant temperature of 195°C in order to prevent photorefractive effects. In either cases, during propagation in the 5mm-long poled region of the sample, the FF beam generated a frequency doubled signal via rQPM [28].

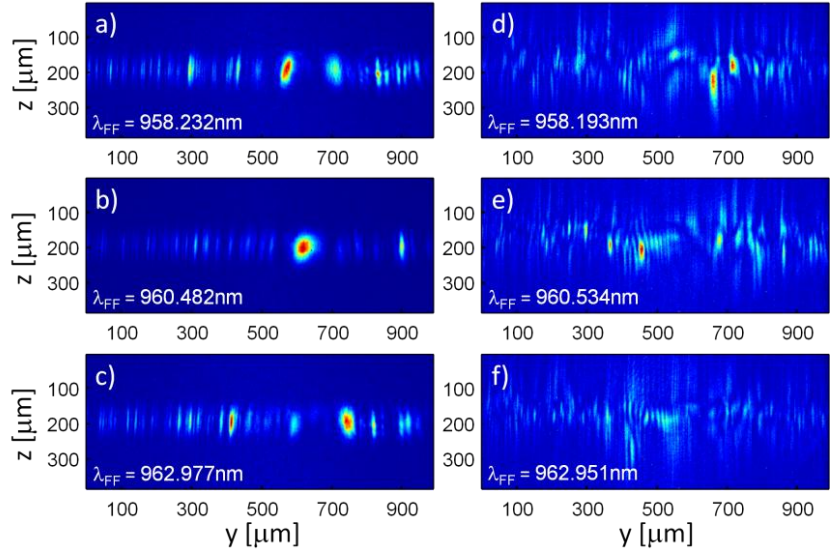

Fig. 2. Acquired far-field SH intensity patterns via rQPM in (a, b, c) bulk and (d, e, f) waveguide samples (the FF wavelength, measured by an optical spectrum analyzer, is indicated in the legends).

In order to acquire FF and SH far-field intensity profiles we positioned a 10x microscope objective by the sample output facet, with the latter 1.5mm away from the ideal object distance; we collected the image with a high resolution CCD camera. A distance > 1.5mm between lens and field profiles at the sample exit is large enough for the far-field condition to hold. In bulk measurements, as previously reported in Ref [28], the FF beam preserved its Gaussian shape during and after propagation, while the generated SH exhibited a set of finger-like spots, parallel to one another and to the z axis of the crystal, with intensities and distribution strongly depending on location and wavelength of the FF input (Fig. 2(a,b,c,)). During SHG via rQPM in slab waveguides, the symmetry described above for the bulk is broken because of the simultaneous presence of two modes at the SH. In particular, random features in the y-coordinate witness the random MTSR of the poled region (as in bulk), while intensity discontinuities along z are due to the interference between the generated $TM_0$ and $TM_1$ modes; since the latter exhibit different effective indices and transverse distributions, their overlap in the far field gives rise to z-modulated distributions for each input (FF) wavelength. The combination of random features along y and far-field modal interference along z results in the complex patterns displayed in Fig. 2(d, e, f).

## 3. Data analysis

rQPM (i.e. QPM with randomness in MTSR) contributes several reciprocal lattice vectors in the Fourier spatial domain; hence, SH and FF are non-collinear either in bulk or in planar waveguide configurations. Using the far-field images we linked each SH peak with the angle φ between the wavevectors $\mathbf{k}_\omega$ and $\mathbf{k}_{2\omega}$ at FF and SH, respectively. QPM occurs at first-order via the grating vector $\mathbf{k}_G$, with $|\mathbf{k}_G| = 2\pi/\Lambda_G$ and $\Lambda_G$ the average period of the random nonlinear coefficient along a specific direction of propagating SH (see Fig. 3).

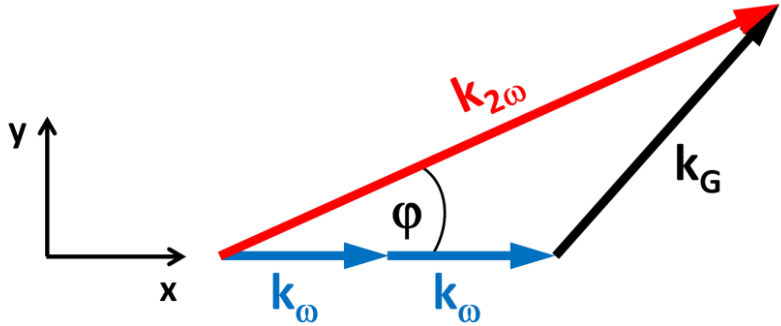

Fig. 3. Sketch of the momentum conservation for SHG via rQPM.

Therefore, three-wave mixing momentum conservation (i.e. phase matching) $\mathbf{k}_{2\omega} - 2\mathbf{k}_\omega - \mathbf{k}_G = \mathbf{0}$ allows using the experimental data (peak positions, i.e., angles φ) to estimate the average period through the equation:

$$k_{2\omega}\cos\varphi - 2k_\omega - k_G\sqrt{1-\left(\frac{k_{2\omega}}{k_G}\sin\varphi\right)^2} = \frac{4\pi}{\lambda_{FF}}n_2\cos\varphi - \frac{4\pi}{\lambda_{FF}}n_1 - \frac{2\pi}{\Lambda_G}\sqrt{1-\left(2\frac{\Lambda_G}{\lambda_{FF}}n_2\sin\varphi\right)^2} = 0 \quad (1)$$

where $\lambda_{FF}$ is the FF wavelength, $n_1$ and $n_2$ are the bulk (waveguide) LT refractive (effective) indices for the extraordinary (TM) polarization with electric field along z at FF and SH, respectively. In the guided-wave case, in particular, while $n_1$ is the effective index of the $TM_0^{FF}$ mode at FF, $n_2$ may refer to either $TM_0^{SH}$ or $TM_1^{SH}$ modes at SH. Bulk refractive indices at FF and SH were calculated from the Sellmeier equations for LT [41], while the effective indices at the two wavelengths were evaluated with a modal solver and the appropriate graded-index profile of the PE waveguide.

We numerically solved Eq. (1) for the average period $\Lambda_G$, using all data on angles φ for about 60 distinct input FF wavelengths between 957 and 963 nm in either bulk or waveguide samples. The resulting (unweighted) histogram provides information on the frequency of occurrence of each periodicity; the same information can be conveniently weighted by the SH conversion efficiency along the directions corresponding to the given periods. The presence of two TM guided modes at SH was exploited to improve the statistical analysis, because two different intensity distributions can be examined at each wavelength, thus roughly doubling the number of estimates.

Figures 4 and 5 display raw and weighted results obtained for bulk and surface samples, respectively. Because of the narrow interval of FF wavelengths used in our experimental campaign we were able to access only the range of periodicities between 3.3 and 5.5 μm. This is due to the connection between the addressable range of spatial periodicities and the tuning range of the FF laser used for the generation of the forward-propagating SH, as previously discussed in Ref [29]. A period of 5.4μm corresponds to collinear QPM SHG for $\lambda_{FF}$ = 960nm. The found bulk periodicity is consistent with what previously estimated in a similar LT sample [29].

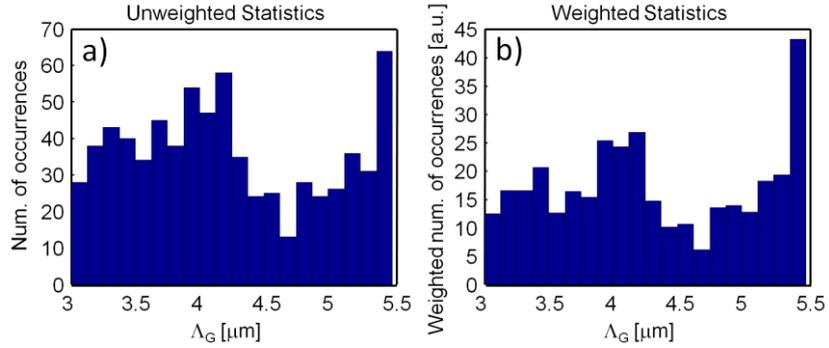

Fig. 4. (a) Histogram with the number of occurrences of values of $\Lambda_G$ between 3.0 and 5.5μm when the input FF propagates in the bulk. (b) Same as in (a) but with number of occurrences weighted by SH conversion efficiency along the direction corresponding to a given period.

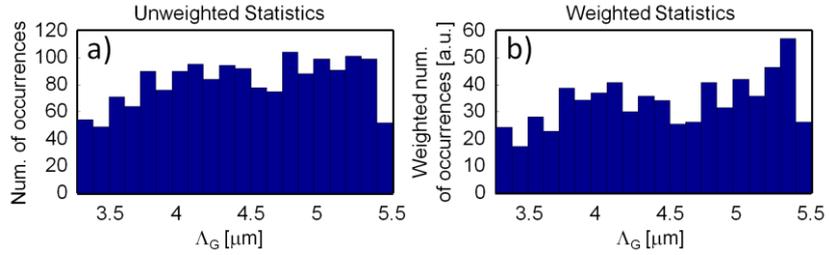

Fig. 5. (a) Histograms with the number of occurrences of $\Lambda_G$ between 3.0 and 5.5μm in the slab waveguide. (b) Same as in (a) but with the number of occurrences is weighted with SH conversion efficiency along the direction corresponding to a given period.

## 4. Conclusions

In conclusion we investigated SHG via rQPM both in bulk LT and in PE waveguides. Measurements of SH intensity profiles in both cases were acquired and employed as a non-invasive tool for estimating the average QPM periodicity in the x-y distribution of the nonlinear coefficient. The presence of a surface guiding structure allowed us to obtain this information both from the bulk and the PE surface region. This statistical approach is a simple and non-invasive means for evaluating the distribution of inverted ferroelectric domains in poled ferroelectric crystals for QPM parametric mixing.

## Acknowledgments

This work was supported by the Italian MIUR through PRIN 2007 no. 2007CT355.